\documentclass[preprintnumbers,twocolumn,prl,aps,superscriptaddress,footinbib,amsfonts,amsmath,amssymb,showpacs]{revtex4-1}

\usepackage[T1]{fontenc}
\usepackage{blindtext}

\usepackage{graphics,graphicx}
\usepackage{dcolumn}
\usepackage{bm}
\usepackage{epsfig}
\usepackage[usenames]{color}
\usepackage[hidelinks]{hyperref} 
\usepackage{mathbbol}
\usepackage{epstopdf}
\usepackage{simplewick}
\usepackage[utf8]{inputenc} 
\usepackage{slashed}
\usepackage{soul}
\usepackage[dvipsnames]{xcolor}
\usepackage[export]{adjustbox}

\usepackage{multirow}

\usepackage{lineno} 

\usepackage{hyperref} 
\hypersetup{
    colorlinks,
    citecolor=black,
    filecolor=black,
    linkcolor=black,
    urlcolor=black
}

\newcommand{\nn}{\nonumber}

\renewcommand{\(}{\left(}
\renewcommand{\)}{\right)}

\begin{document}


\title{Determination of Collins-Soper kernel from cross-sections ratios}

\newcommand*{\DESY}{Deutsches Elektronen-Synchrotron DESY, Germany}\affiliation{\DESY}
\newcommand*{\REG}{Institut f\"ur Theoretische Physik, Universit\"at Regensburg, D-93040 Regensburg, Germany}\affiliation{\REG}
\newcommand*{\MAD}{Dpto. de F\'isica Te\'orica \& IPARCOS, Universidad Complutense de Madrid, E-28040 Madrid, Spain}\affiliation{\MAD}

\author{Armando~Bermudez~Martinez}\email{armando.bermudez.martinez@desy.de}\affiliation{\DESY}
\author{Alexey~Vladimirov}\email{alexeyvl@ucm.es}\affiliation{\REG}\affiliation{\MAD}

\begin{abstract}
\noindent
We present a novel method of extraction of the Collins-Soper kernel directly from the comparison of differential cross-sections measured at different energies. Using this method, we analyze the pseudo-data generated by the CASCADE event generator and extract the Collins-Soper kernel predicted by the parton-branching model in the wide range of transverse distances. The procedure can be applied, with minor modifications, to the real measured data for Drell-Yan and SIDIS processes.
\end{abstract}

\maketitle

{\bf Introduction.} The primary goal of the modern Quantum-Chromo Dynamics (QCD) is to understand the inner structure of the nucleon and the forces that bind its constituents together. The confinement mechanism prevents any direct exploration of the hadron's insides, and thus one should employ indirect approaches. The main of these approaches is the analysis of the differential cross-sections of particle scattering. The major tool for interpreting scattering cross-sections is the factorization theorems \cite{Collins:1989gx}, which are formulated in terms of universal parton distributions, each of which highlights a specific aspect of parton's dynamics. Among a variety of parton distributions the special role is played by the Collins-Soper (CS) kernel \cite{Collins:1984kg}, which emerges from the factorization theorems for the transverse-momentum-differential cross-sections \cite{Collins:2011zzd, Echevarria:2011epo, Becher:2010tm}.  

Despite being a part of the factorization theorem, the CS kernel is conceptually different from other distributions. First and foremost, the CS kernel is not a characteristic of a hadron. It provides us with information about the long-range forces acting on quarks that are imposed solely by the non-trivial structure of the QCD vacuum \cite{Vladimirov:2020umg}. In that sense, the CS kernel is the most fundamental distribution in the modern framework of factorization theorems, and for that reason, the determination of the CS kernel has been performed in many works.

\begin{figure}[t]
\begin{center}
\includegraphics[width=0.45\textwidth]{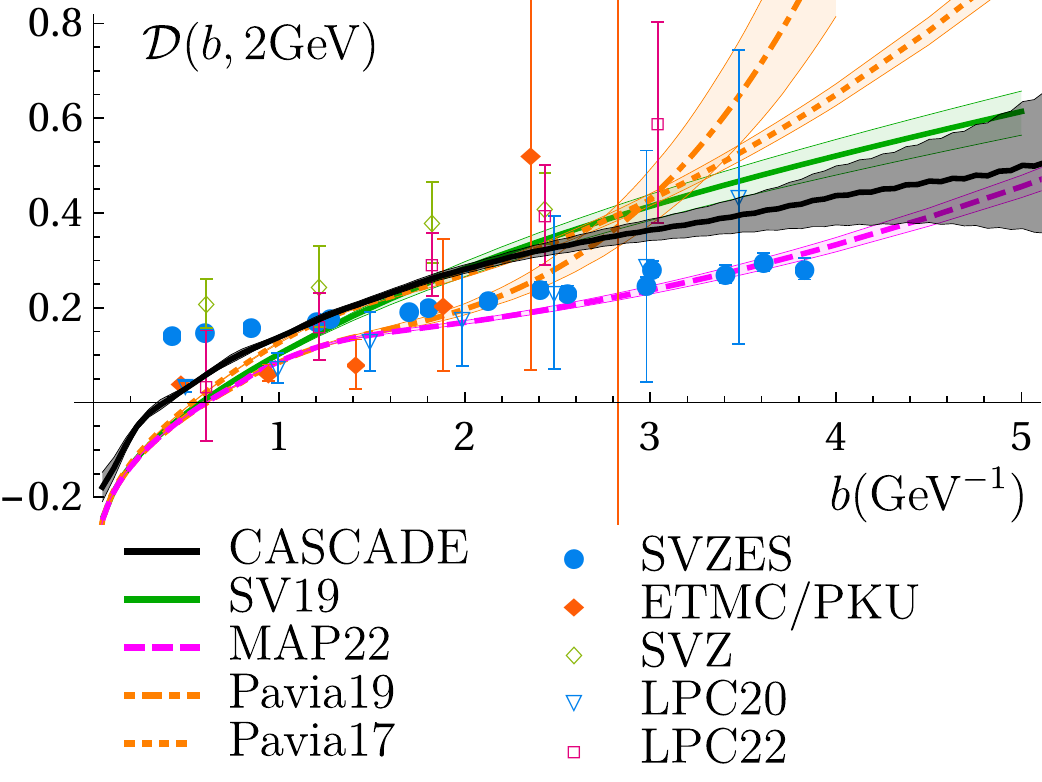}
\end{center}
\caption{\label{fig:RAD-final} Comparison of the CS kernels obtained in different approaches. CASCADE curve is obtained in this work. The curves SV19, MAP22, Pavia19 and Pavia17 are obtained from the fits of Drell-Yan and SIDIS data in refs. \cite{MAP22}, \cite{Scimemi:2019cmh}, \cite{Bacchetta:2019sam}, and \cite{Bacchetta:2017gcc}, correspondingly. Dots represent the computations of CS kernel on the lattice, with SVZES, ETMC/PKU, SVZ, LPC20 and LPC22 corresponding to refs.\cite{Schlemmer:2021aij}, \cite{Li:2021wvl}, \cite{Shanahan:2021tst}, \cite{LatticeParton:2020uhz}, and \cite{LPC:2022ibr}.}
\end{figure}

At present, there are two approaches to the determination of the CS kernel. The most traditional one is the extraction of the CS kernel from the experimental data for TMD cross-sections. The latest extractions are based on the global fits of Drell-Yan and Semi-Inclusive Deep-Inelastic scattering processes \cite{Bacchetta:2017gcc, Scimemi:2017etj, Bertone:2019nxa, Scimemi:2019cmh, Bacchetta:2019sam}. Another approach uses the QCD lattice simulations. It has been suggested recently in refs.\cite{Ebert:2018gzl, Ji:2019ewn, Vladimirov:2020ofp}, and the results of the first simulations are already available \cite{Shanahan:2020zxr, Schlemmer:2021aij, Shanahan:2021tst, Chu:2022mxh}. However, so far, the different extractions do not demonstrate a good agreement (see the comparison in ref.\cite{Vladimirov:2020umg}, and also in fig.\ref{fig:RAD-final}), which is due to the large systematic uncertainties. The phenomenological extractions are biased by the fitting ansatz, while the lattice extractions are contaminated by large power corrections. In this letter, we suggest a direct way of extracting the CS kernel from the scattering data. The method inherits the main idea of lattice computations, namely, to form the proper ratios of observables. In this work, we consider cross-section ratios.

To demonstrate the method's power, we use pseudo data generated by the CASCADE event generator \cite{Baranov:2021uol}. Predictions from CASCADE rely on the Parton Branching (PB) method \cite{Hautmann:2017xtx, Hautmann:2017fcj} for the evolution of transverse momentum dependent (TMD) parton densities \cite{BermudezMartinez:2018fsv} and provide an excellent description of Drell-Yan transverse momentum spectrum measurements in a wide range of Drell-Yan mass and center-of-mass energies \cite{BermudezMartinez:2020tys, Martinez:2021mzy, BermudezMartinez:2019anj, CMS:2019raw, Martinez:2021chk}. This makes the PB-TMD simulations by CASCADE a solid ground for applying the procedure proposed hereby.

{\bf Method.} The method is founded on the leading power transverse momentum dependent (TMD) factorization theorem \cite{Collins:2011zzd, Echevarria:2011epo}. For the concreteness, we consider the Drell-Yan pair production process $h_1+h_2\to \gamma^*(\to \ell^+\ell^-)+X$. Other processes, for which the TMD factorization is formulated, can be analyzed in an analogous way. The TMD factorization theorem for the unpolarized Drell-Yan process reads \cite{Collins:1989gx, Collins:2011zzd, Echevarria:2011epo, Becher:2010tm}
\begin{eqnarray}\nn
\frac{d\sigma}{dQ^2 dy d q_T^2}=\frac{2\pi}{9}\frac{\alpha_{\text{em}}^2(Q)}{s Q^2}|C_V(Q,\mu_Q)|^2\int_0^\infty db b J_0(bq_T)
\\\label{TMD-fack} \times \sum_q e^2_q f_{q,h_1}(x_1,b;\mu_Q,Q^2)f_{\bar q,h_2}(x_2,b;\mu_Q,Q^2),~~~
\end{eqnarray}
where $Q$, $y$ and $q_T$ are the invariant mass, rapidity and transverse momentum of the virtual photon, $\mu_Q\sim Q$ is the factorization scale, $s$ is the invariant mass of the initial state, $e_q$ are the electric charges of quarks, $\alpha_{\text{em}}$ is the fine-structure constant and $J_0$ is the Bessel function of the first kind. The variables $x_1$ and $x_2$ are longitudinal momentum fractions
\begin{eqnarray}
x_1=\frac{Q}{\sqrt{s}}e^{y},\qquad x_2=\frac{Q}{\sqrt{s}}e^{-y}.
\end{eqnarray}
The hard coefficient function $C_V$ is entirely perturbative and known up to next-to-next-to-next-to-leading order (N$^3$LO) \cite{Gehrmann:2010ue}. The functions $f$ are non-perturbative unpolarized TMD distributions.

The CS kernel is hidden in the $Q$-dependence of TMD distributions that is described by a pair of evolution equations \cite{Aybat:2011zv, Chiu:2012ir, Scimemi:2018xaf}
\begin{eqnarray}\label{evol}
\frac{d f_{q,h}(x,b;\mu,\zeta)}{d\ln \mu^2}=\frac{\gamma_F(\mu,\zeta)}{2}f_{q,h}(x,b;\mu,\zeta),
\\\nn
\frac{d f_{q,h}(x,b;\mu,\zeta)}{d\ln \zeta}=-\mathcal{D}(b,\mu)f_{q,h}(x,b;\mu,\zeta).
\end{eqnarray}
Here, $\gamma_F$ is the TMD anomalous dimension which is perturbative and known up N$^3$LO \cite{Gehrmann:2010ue}, and $\mathcal{D}$ is the non-perturbative CS kernel \footnote{There is no common notation for CS kernel. In the literature, it is also denoted as $\tilde K=\gamma_\nu^f=-F_{f\bar f}=-2\mathcal{D}$  \cite{Becher:2010tm, Aybat:2011zv, Chiu:2012ir, Scimemi:2018xaf}.}. Thus, to extract the CS kernel one must explore the $Q$-dependence of the cross-section.

The essential complication of any phenomenological analysis with the TMD factorization is that all non-perturbative functions are defined in the position space. We perform the inverse Hankel transform of the cross-section
\begin{eqnarray}\label{def:Sigma}
\Sigma(s,y,Q,b)=\int_0^\infty dq_T\,q_TJ_0(q_T b)\frac{d\sigma}{dQ^2 dy d q_T^2}.
\end{eqnarray}
The formula (\ref{TMD-fack}) is valid at small-$q_T/Q$, and the corrections to it are estimated as $\sim1\%$ at $q_T=0.1Q$ \cite{Scimemi:2019cmh}. Consequently, $\Sigma$ is accurately (up to $1\%$) described in the terms of TMD distributions for $b\gtrsim (0.1 Q)^{-1}$. 

The main idea of the method is to compare $\Sigma$'s measured at different $Q$'s ($Q_1$ and $Q_2$), such that the TMD distributions $f$ exactly cancel in the ratio. To perform the cancellation we adjust the values of $s$ such that the variables $x_{1,2}$ are identical. We compute
\begin{eqnarray}\label{main-ratio}
\frac{\Sigma(s_1,y,Q_1,b)}{\Sigma(s_2,y,Q_2,b)}
=\(\frac{Q_2}{Q_1}\)^4 Z(Q_1,Q_2)e^{2\Delta(b;Q_1\to Q_2)},
\end{eqnarray}
where $s_1/s_2=Q_1^2/Q_2^2$. The function $Z$ is entirely perturbative
\begin{eqnarray}\label{def:Z}
Z(Q_1,Q_2)=\frac{\alpha_{\text{em}}^2(Q_1)|C_V(Q_1,\mu_{Q_1})|^2}{\alpha_{\text{em}}^2(Q_2)|C_V(Q_2,\mu_{Q_2})|^2}.
\end{eqnarray}
The function $\Delta$ is resulted from the evolution of TMD distribution to the same scale by equations (\ref{evol}),
\begin{eqnarray}\label{evolution-integral}
\Delta(b,Q_1\to Q_2)&=&\int_P\(\gamma_F(\mu,\zeta)\frac{d\mu}{\mu}-\mathcal{D}(b,\mu)\frac{d\zeta}{\zeta}\),
\end{eqnarray}
where $P$ is a path connecting points $(\mu_{Q_1},Q_1^2)$ and $(\mu_{Q_2},Q_2^2)$ in the $(\mu,\zeta)$-plane \cite{Scimemi:2018xaf}. Thus, the only non-perturbative content in the formula (\ref{main-ratio}) is the CS kernel.

To invert the formula (\ref{main-ratio}) we use the rectangular contour for the integration path in eqn. (\ref{evolution-integral}) and find
\begin{eqnarray}\label{D-main}
&&\mathcal{D}(b,\mu_0)=
\\\nn 
&&\frac{\ln\(\frac{\Sigma_1}{\Sigma_2}\)-\ln Z(Q_1,Q_2)-2 \Delta_R(Q_1,Q_2;\mu_0)}{4\ln(Q_2/Q_1)}-1,
\end{eqnarray}
where $\Sigma_1/\Sigma_2$ is shorthand notation for the ratio (\ref{main-ratio}), and 
\begin{eqnarray}\label{def:DeltaR}
\Delta_R(Q_1,Q_2;\mu_0)=\int_{\mu_{Q_2}}^{\mu_{Q_1}}\frac{d\mu}{\mu} \gamma_F(\mu,Q_1)\qquad\qquad
\\ \nn  -2\ln\(\frac{Q_1}{Q_2}\)\int_{\mu_0}^{\mu_{Q_2}}\frac{d\mu}{\mu}\Gamma_{\text{cusp}}(\mu),
\end{eqnarray}
with $\Gamma_{\text{cusp}}$ being the cusp anomalous dimension. The last term in eqn.(\ref{def:DeltaR}) evolves CS kernel to the scale $\mu_0$, which is used to compare different extractions. Apart of the ratio $\Sigma_1/\Sigma_2$ all terms in equation (\ref{D-main}) are perturbative, and nowadays know up to N$^3$LO. Therefore, the formula (\ref{D-main}) can be used to extract CS kernel directly from the data without any further approximation.

\begin{table}[b]
\begin{tabular}{c||c|c|c}
Init.state & $Q$ [GeV] & $\sqrt{s}$ [GeV] & $\delta y$
\\\hline
\multirow{4}{*}{pp} & 12 & 655.2 &  \multirow{4}{*}{4}
\\ & 16 & 873.6 &
\\ & 20 & 1092. &
\\ & 24 & 1310. &
\\\hline
\multirow{2}{*}{pp} & 12 & 88.7 &  \multirow{2}{*}{2}
\\ & 16 & 118.2 &
\\\hline
\multirow{2}{*}{pp} & 12 & 241.0 &  \multirow{2}{*}{3}
\\ & 16 & 321.4 &
\\\hline
\multirow{2}{*}{pp} & 12 & 1781. &  \multirow{2}{*}{5}
\\ & 16 & 2375. &
\\\hline
\multirow{3}{*}{p$\bar{\text{p}}$} & 12 & 655.2 &  \multirow{3}{*}{4}
\\ & 16 & 873.6 &
\\ & 20 & 1092. &
\end{tabular}
\caption{\label{tab:runs} Parameters of the generated events. For each case $\delta Q=5\% \cdot Q$.}
\end{table}

Practically, the experimental measurements for differential cross-sections are presented by a collection of points in bins of $(Q,y,q_T)$. Therefore, the transformation (\ref{def:Sigma}) cannot be computed analytically but by the discrete Hankel transform. Herewith, one should find a balance between the statistical precision of $d\Sigma$ (which usually decreases at low-$q_T$) and the range of $b$ (larger $b$ requires lower $q_T$). Alternatively, the experimental curve can be fit by an analytical form, and the transformation (\ref{def:Sigma}) is performed analytically. This path, however, introduces uncertainty due to the curve parameterization. 

The integration over $q_T$ leaves intact the dependence on $Q$ and $y$, which can be used to increase the statistical precision. We introduce
\begin{eqnarray}
\Sigma(s,Q,b)=\int_{Q-\delta Q}^{Q+\delta Q}dQ^2 \int_{-\delta y}^{\delta y}dy\, d\Sigma(s,y,Q,b),
\end{eqnarray}
where $\delta Q$ and $\delta y$ are sizes of the bin. These function can be also used in the ratio $\Sigma_1/\Sigma_2$ with the only restriction that $\delta Q\ll Q$. In this case, the effects of variation of $Q$ within the bin could be neglected. There is no limitation for $\delta y$, except that $\delta y$ is the same for $\Sigma_1$ and $\Sigma_2$. 

The equation (\ref{D-main}) is valid for any kind of initial hadrons. This property can be used to cross-validate the extraction. The values of $Q$ should be large enough to neglect the QCD power and target-mass corrections $Q\gg \Lambda_{\text{QCD}}$, and small enough to neglect the contribution of the $Z$-boson intermediate state $Q\ll M_Z$. It provides a very large window of available energies. The corrections for the $Z$-boson can be included in eqn.(\ref{D-main}) by modifying expression for $Z$ eqn.(\ref{def:Z}).

{\bf Extraction of the CS kernel using CASCADE.} To test the proposed approach, we study the pseudo-data generated with the CASCADE event generator \cite{Baranov:2021uol}. The parameters for the evolution of TMD parton densities \cite{BermudezMartinez:2018fsv} were determined solely from inclusive HERA deep inelastic scattering measurements. The predictions provided by CASCADE describe the Drell-Yan transverse momentum spectrum for both low-energy measurements from the PHENIX, NUSEA, R209, E605 experiments \cite{BermudezMartinez:2020tys, Martinez:2021mzy}, as well as high-energy data from the LHC experiments CMS, and ATLAS \cite{BermudezMartinez:2019anj, CMS:2019raw, Martinez:2021chk}. Importantly, the PB algorithm does not explicitly employ the TMD factorization theorem. In particular, there are no parameters specially dedicated to the CS kernel, and thus the CS kernel emerges via the interplay of the parameters controlling the PB-TMD evolution in the CASCADE generator. Therefore, our determination of the CS kernel is also an ultimate test of compatibility between the PB and the factorization approaches.

We have generated several sets of pseudo-data. These sets include a variety of combinations for parameters, namely, different $Q$, $\delta y$, and hadron types. The CS kernel is independent of these parameters and thus, a priory all combinations should result in the same value. Their kinematic setups are shown in table \ref{tab:runs}.

\begin{figure}[t]
\begin{center}
\includegraphics[width=0.4\textwidth]{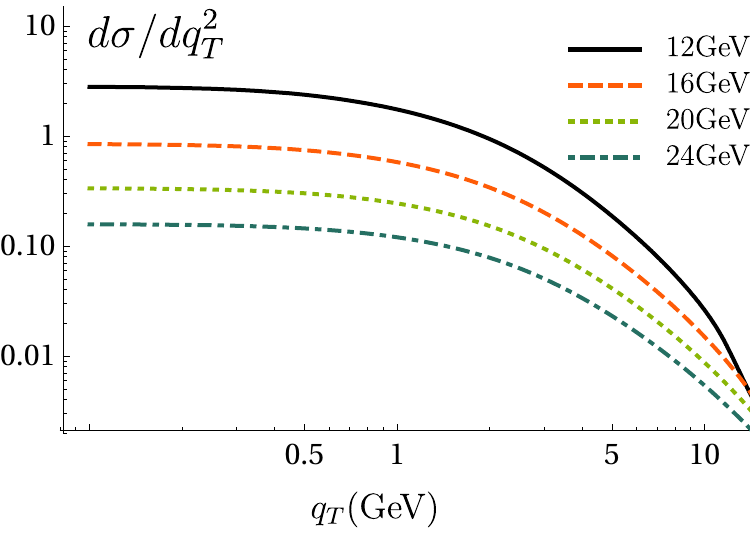}
\\
\includegraphics[width=0.4\textwidth]{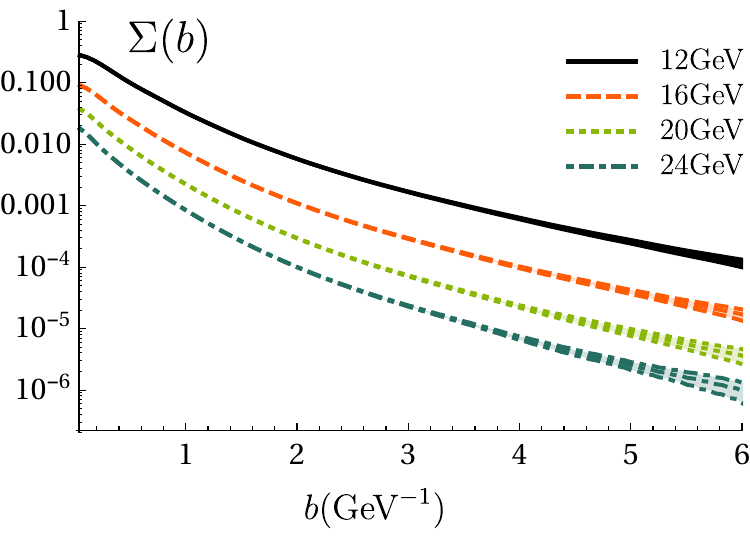}
\end{center}
\caption{\label{fig:DY} Example of pseudo-data in the momentum (top plot) and position (bottom plot) spaces. The statistical uncertainty is shown by the width of the line. The shown cases correspond to $pp$-scattering integrated in $|y|<4$.}
\end{figure}

The inverse Hankel transform has been performed using the algorithm \cite{doi:10.1063/1.468428}. The algorithm expects that the input function vanishes beyond the presented domain. It is a good approximation since the cross-section for the Drell-Yan process drops rapidly at large-$q_T$. For the considered cases, the relative numerical uncertainty due to the truncation is $10^{-5}-10^{-6}$ and can be safely neglected. 

The effective range and accuracy of the discrete Hankel transform depend on the density and range of the input cross-section. So, to obtain a stable curve at the large-b, one needs a large number of points at small-$q_T$. In particular, we collect events into narrow $q_T$-bins with the size $0.05$GeV, which allows us to reach $b\sim 4-5$GeV$^{-1}$. At larger values of $b$, the inverse function is sensitive to the finite-bin effects and becomes unstable. The examples of the cross-section in momentum and position spaces are shown in fig.\ref{fig:DY}. We employ the bootstrap method to estimate the propagation of statistical uncertainty from the momentum to the position space. During the re-sampling, we also vary the central value of $q_T$ within a bin, which allows us to estimate the uncertainty due to finite bin size at large-$b$. 

\begin{figure}[t]
\begin{center}
\includegraphics[width=0.45\textwidth]{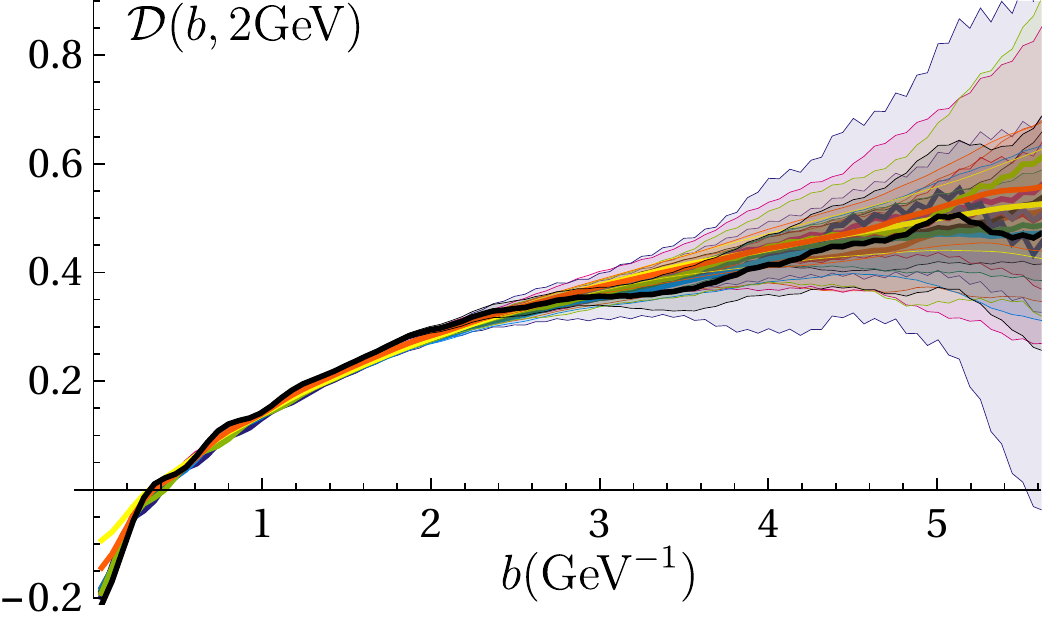}
\\
\includegraphics[width=0.45\textwidth]{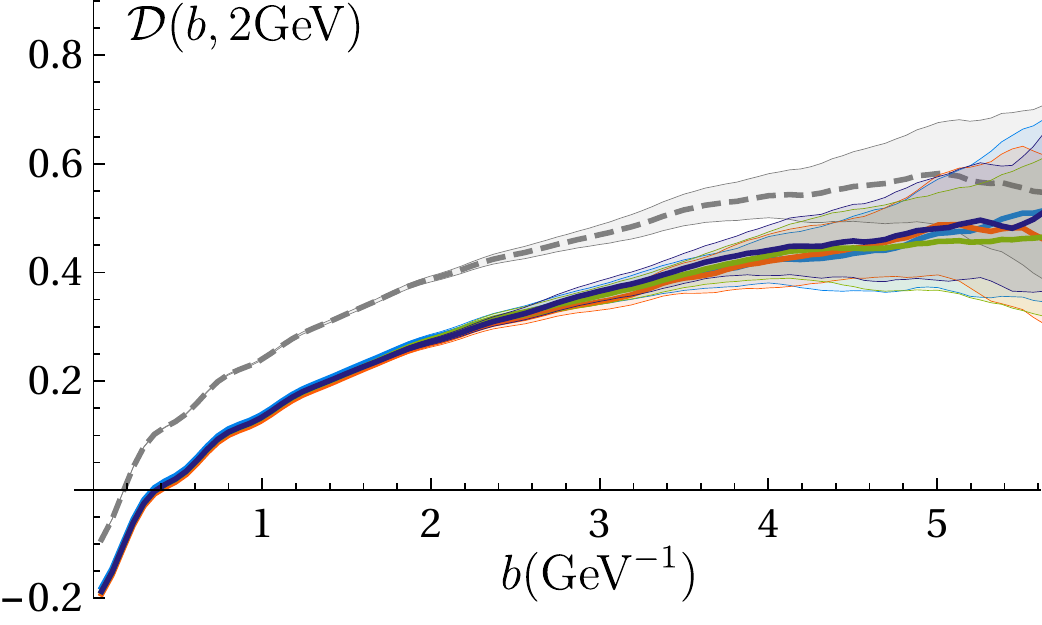}
\end{center}
\caption{\label{fig:CS-different} Comparison of CS kernels extracted from different combinations of the pseudo-data. The top plot shows all possible (twelve) combinations of pseudo-data with different kinematics, listed in the table \ref{tab:runs}. The bottom plot show extractions made with different input collinear PDFs. The solid lines are the central values. The shaded areas are the statistical uncertainty. The oscillations at $b\sim 4-6$GeV$^{-1}$ are due to the finite bin size in the $q_T$-space. The gray dashed line in the lower plot shows the effect of incomplete cancellation of parton's momentum if PDFs in the comparing cross-section are different (here, CT18 vs. CASCADE).}
\end{figure}

Using the sets listed in table \ref{tab:runs}, we compose twelve ratios $\Sigma_1/\Sigma_2$. Each ratio provides an independent value of CS kernel (\ref{D-main}). The collection of resulting curves at $\mu=2$GeV is shown in fig.\ref{fig:CS-different}(top). Apparently all curves are in a perfect agreement for $b>0.4$GeV$^{-1}$. It manifests that the CASCADE event generator supports the TMD factorization theorem. Note that three extractions are made with proton-antiproton collision, and they are indistinguishable from the proton-proton cases. It confirms the universality of the CS kernel.

For an extra test of the universality, we have simulated the pseudo-data ($pp$ collisions at $Q=12$ and $16$ GeV, $\delta y=4$) using different collinear PDFs. Namely, the CT18 \cite{Hou:2019efy}, MSHT20 \cite{Bailey:2020ooq} and NNPDF3.1 \cite{NNPDF:2017mvq}. The comparison of these cases is shown in fig.\ref{fig:CS-different}(top). These curves are also in the perfect agreement, which shows that the TMD evolution within CASCADE is independent of the DGLAP evolution. For a demonstration of a possible misbehavior, we show (in the gray-dashed line in fig.\ref{fig:CS-different}(bottom)) the CS kernel extracted from the ratio with different collinear PDFs (CASCADE at $Q=12$GeV to CT18 at $Q=16$GeV), where the parton distributions do not cancel exactly.

All extractions are made using $\mu_Q=Q$, and N$^3$LO perturbative accuracy for functions $Z$ and $\Delta_R$. The perturbative expansion is very stable, which we test by varying the scale $\mu_Q\in[Q/2,2Q]$. The size of the scale variation band is constant in $b$. The maximum variation among all extractions is $(-0.0016,+0.0008)$ in the absolute value, which is smaller than the statistical uncertainty.

The final curve for the CS kernel predicted by the CASCADE event generator is obtained by combining all twelve extractions. The comparison with other extractions of CS kernel is shown in fig.\ref{fig:RAD-final}. The CASCADE extraction lightly disagrees with the perturbative curve ($b<1$GeV$^{-1}$), but in agreement with the SV19 \cite{Scimemi:2019cmh} and Pavia17 \cite{Bacchetta:2017gcc} for $1<b<3$GeV$^{-1}$. 

The fit of the large-$b$ part by a polynomial gives
\begin{eqnarray}
\mathcal{D}(b,\mu)\sim [(0.069\pm 0.031)\text{GeV}]\times b,
\end{eqnarray}
with a negligible quadratic part. We conclude that the CASCADE suggests a linear asymptotic, which was also used in the SV19 series of fits \cite{Bertone:2019nxa, Scimemi:2019cmh, Bury:2022czx}, and supported by theoretical estimations \cite{Vladimirov:2020ofp, Schweitzer:2012hh}

{\bf Conclusions.} We have presented the method of direct extraction of the CS kernel from the data, using the proper combination of cross-sections with different kinematics. For explicitness, we considered the case of the Drell-Yan process, but the method can be easily generalized to other processes such as SIDIS, semi-inclusive annihilation, Z/W-boson production, and their polarized versions. 

The method is tested using the pseudo-data generated by the CASCADE event generator, and the corresponding CS kernel is extracted. Amazingly, all expected properties of the CS kernel (such as universality) are observed in the CASCADE generator. This non-trivially supports both the TMD factorization and the PB approaches and solves an old-stated problem of comparison between non-perturbative distributions extracted within these approaches \cite{Abdulov:2021ivr, Angeles-Martinez:2015sea}.

The procedure can be applied to the real experimental data without modifications. In this case, the uncertainties of extraction will be dominated by the statistical uncertainties of measurements since many systematic uncertainties cancel in the ratio. Thus the method is feasible for modern and future experiments, such JLab \cite{Dudek:2012vr, Accardi:2020swt}, LHC \cite{Amoroso:2022eow}, and EIC \cite{AbdulKhalek:2021gbh, Khalek:2022bzd}. They can be applied to already collected data after a rebinning. Importantly, the procedure is model-independent and provides access to the CS kernel based on the first principles.

{\bf Acknowledgments.}
We thank Hannes Jung and Francesco Hautmann for discussions, and also Qi-An Zhang and Alessandro Bacchetta for providing us with their extractions. A.V. is funded by the \textit{Atracci\'on de Talento Investigador} program of the Comunidad de Madrid (Spain) No. 2020-T1/TIC-20204. This work was partially supported by DFG FOR 2926 ``Next Generation pQCD for  Hadron  Structure:  Preparing  for  the  EIC'',  project number 430824754

\bibliography{biblio}

\end{document}